\documentclass[aps,prl,reprint,showpacs,superscriptaddress,floatfix]{revtex4-2}
\usepackage[utf8]{inputenc}
\usepackage{graphicx}
\usepackage{booktabs}
\usepackage{tabularx}
\usepackage{hyperref}
\usepackage{orcidlink}
\usepackage{amsmath}
\hypersetup{
colorlinks=true,final=true,
        linkcolor=blue,
        citecolor=blue,
        filecolor=blue,
        urlcolor=blue,
}

\usepackage{changes}

\begin{document}
\title{Squeezing dynamical singlets in bilayer nickelates
}
\author{Harrison LaBollita\,\orcidlink{0000-0002-6699-8577}}
\email{hlabollita@flatironinstitute.org}
\affiliation{Center for Computational Quantum Physics, Flatiron Institute, New York, New York 10010, USA}
\author{Andrew J. Millis\,\orcidlink{0000-0001-9413-6344}}
\affiliation{Center for Computational Quantum Physics, Flatiron Institute, New York, New York 10010, USA}
\affiliation{Department of Physics, Columbia University, New York, New York 10027}
\author{Olivier~Gingras\,\orcidlink{0000-0003-3970-6273}}
\affiliation{Center for Computational Quantum Physics, Flatiron Institute, New York, New York 10010, USA}
\affiliation{Université Paris-Saclay, CNRS, CEA, Institut de physique théorique, 91191, Gif-sur-Yvette, France}

\begin{abstract}
    We present realistic calculations within the density functional plus cluster dynamical mean-field formalism indicating that the physics of the the bilayer Ruddlesden-Popper  nickelates is to a significant extent controlled by interlayer ``dynamical singlets'' which are formed from the $3z^{2}-r^{2}$ orbitals singly occupied by electrons and are hybridized with itinerant planar $x^{2}-y^{2}$ orbitals. The hybridization is found to respond differently to hydrostatic pressure and to epitaxial strain, capturing the experimentally observed dichotomy between bulk single crystals and epitaxial thin films and reproducing several experimental results including angle-resolved photoemission and transport measurements. 
\end{abstract}

\maketitle


\textit{Introduction.---}The bilayer Ruddlesden-Popper (RP) nickelates have recently emerged as fascinating systems exhibiting a range of correlated electron behavior, and with systematic differences in the evolution of properties under hydrostatic pressure and compressive epitaxial strain~\cite{sun2023superconductivity, Li2026Bulk, tarn2026, ko2025, li2026_enhanced,Zhou2026Superconductivity,Hou2023Emergence,Zhou2025Ambient-pressure,Wang2024Bulk,wang2024_327scpress}. In contrast to the long-known high-critical temperature ($T_c$) cuprates ~\cite{Bednorz1986possible} and the recently studied square-planar nickelates,  where the physics is widely believed to be captured by a single band per plane derived from a transition metal $d^{9}$ atomic configuration~\cite{Li2019superconductivity, Lee2023linear, Pan2021superconductivity, pan2026, Botana2020similarities, Karp2020manybody, Labollita2021electronic, labollita02022correlated, Labollita2022manybody, Lechermann2020late, Lechermann2020multiorbital, lechermann2022, Wang2020hunds, petocchi2020}, the RP compounds are nominally $d^{7+\delta}$ and intrinsically multi-orbital. Understanding the new physics of these multi-orbital systems and its relation to superconductivity and other correlated electron behavior is an important challenge to condensed matter physics.

\begin{figure*}
    \centering
    \includegraphics[width=2\columnwidth]{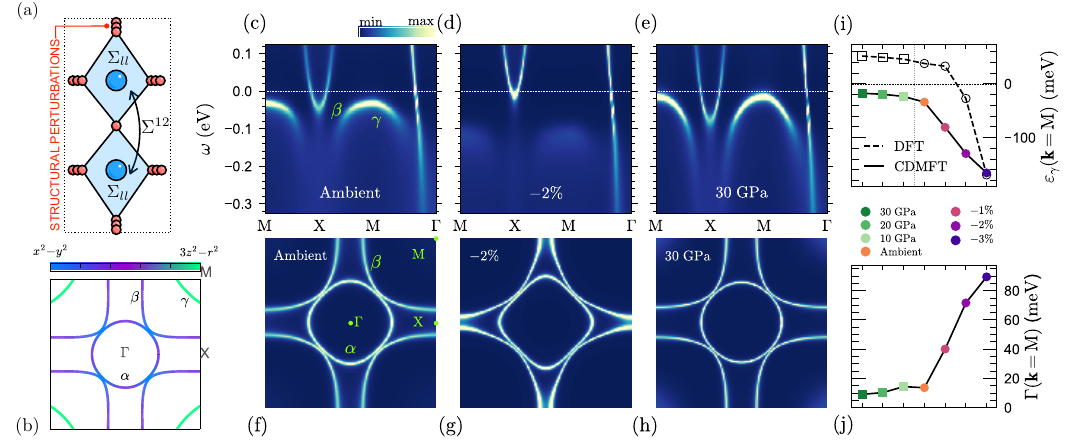}
    \caption{Low-energy many-body electronic structure of La$_{3}$Ni$_{2}$O$_{7}$ under hydrostatic pressure and compressive epitaxial strain.
    (a)~Schematic of the bilayer cluster quantum impurity used in CDMFT, indicating local ($\Sigma_{ll}$) and interlayer ($\Sigma_{12}$) self-energies. Ni and O atoms are shown as blue and red spheres, respectively; overlapping oxygen positions indicate the structural distortions induced by the external structural perturbations.
    (b)~Fermi surface ($k_{z}=0$) of the ambient pressure Wannier Hamiltonian with the orbital character projected onto each sheet: $x^{2}-y^{2}$ (blue), $3z^{2}-r^{2}$ (green), and mixed (purple).
    (c-h)~CDMFT electronic structure at $T = 1/40$ ($\sim 290$ K) for (c,f)~ambient pressure, (d,g)~$-2$\% epitaxial strain, and (e,h)~30~GPa hydrostatic pressure.
    (c-e)~Momentum-resolved spectral function $A(\mathbf{k},\omega)$ along the high-symmetry path M-X-M-$\Gamma$.
    (f-h)~Fermi surface in the $k_{z}=0$ plane.
    (i,j)~Evolution of the $z$-derived $\gamma$ band across the different crystal structures: (i)~binding-energy position at $\textbf{k}=$M and (j)~corresponding linewidth (scattering rate $\Gamma$).
    }
    \label{fig:main}
\end{figure*}

Recent experimental and theoretical work~\cite{sun2023superconductivity, pardo2011, lechermann2023electronic, zhang2023,christiansson2023, labollita2024electronic, yang2024_arpes, leonov2023,ouyang2025phase,chen2024nfl,zhang2023electronic,zhang2023trends,zhang2025general, Luo2023}, has established that the low-energy electronic structure of bilayer RP nickelates, La$_{3}$Ni$_{2}$O$_{7}$, may be described in terms of bilayers with an $x^2-y^2$ and a $3z^2-r^2$ orbital centered on the Ni sites of each layer and very weak electronic coupling between bilayers.
The four Ni-$e_{g}$ orbitals form bonding- and antibonding-like states near the Fermi level, with the $3z^{2}-r^{2}$ orbitals strongly hybridized between the two planes in a bilayer via a bridging apical oxygen, while intersite hopping terms lead to a hybridization between $x^2-y^2$ and $3z^2-r^2$ orbitals on nearby sites in the same plane. The chemical composition implies a formal valence of Ni($3d^{7.5}$), so that three electrons occupy these four orbitals, providing a minimal starting point for a theoretical description of the materials. 

An important aspect is that the material properties  evolve differently under hydrostatic pressure and epitaxial compressive strain \cite{zhao2025,bhatt2025resolvingstructuraloriginssuperconductivity}. Additionally, recent theoretical works have highlighted that non-local interlayer correlations within the RP bilayer can significantly reshape the low-energy electronic structure~\cite{ryee2024_quench, liao2024orbitalselectiveelectroncorrelationshightrm}. Thus, a central challenge is to develop an electronic structure framework that consistently describes the normal-state of bilayer RP nickelates across both bulk single crystals and epitaxial thin-films.

In this Letter, we develop a cluster dynamical mean-field theory (CDMFT) for the normal-state electronic structure of bilayer RP nickelates and its evolution under hydrostatic pressure and epitaxial strain, capturing both local and interlayer electronic correlations within the Ni-$e_{g}$ bilayer cluster. Our calculations show how structural perturbations and electronic correlations cooperate to reshape the low-energy electronic structure, strongly renormalizing the interlayer coupling within the $3z^{2}-r^{2}$ orbitals and drive the system to an orbital-selective singlet-paired Mott regime in which interlayer ``dynamical singlets'' constructed from the $3z^{2}-r^{2}$ orbitals  are hybridized with the itinerant planar $x^{2}-y^{2}$ orbitals on each layer. Our interacting theory qualitatively reproduces several experimental observations, including orbital-selective mass renormalizations seen in angle-resolved photoemission spectroscopy (ARPES)~\cite{yang2024_arpes}, offers a normal-state rationale for the observed enhancement of $T_{c}$ under combined compressive strain and hydrostatic pressure~\cite{li2026_enhanced}, captures the contrasting normal-state transport of strained thin-films at ambient pressure~\cite{Hsu2026Fermi,kumar2026nonfermiliquidbehaviorla3ni2o7} and pressurized bulks single crystals~\cite{Zhang2024hightemp} and may provide a different paradigm for correlated electron physics.

\textit{Low-energy models.---}We construct low-energy Wannier Hamiltonians from density functional theory (DFT)  band structures of La$_{3}$Ni$_{2}$O$_{7}$ under epitaxial strain and hydrostatic pressure, retaining the near Fermi level Ni-$e_{g}$ states, namely the in-plane $d_{x^{2}-y^{2}}$ and out-of-plane $d_{3z^{2}-r^{2}}$ orbitals (hereafter denoted $x$ and $z$, respectively). The Wannier Hamiltonians are augmented with local interactions of the Hubbard-Kanamori form on each Ni site, parameterized by $U=4$~eV (intraorbital),  $U^{\prime}=2.4$~eV (interorbital), and $J=0.8$~eV (Hund's coupling), that are in line with previous studies~\cite{christiansson2023correlated,ryee2024_quench}.

Following Ref.~\onlinecite{ryee2024_quench}, for the bilayer system (two orbitals per layer), we define a four-orbital cluster comprising the two Ni sites and their frontier $e_{g}$ orbitals, which serves as the quantum impurity problem [see Fig.~\ref{fig:main}(a)]. 
The resulting quantum impurity is solved using the continuous-time quantum Monte Carlo method in the hybridization expansion~\cite{PhysRevLett.97.076405, RevModPhys.83.349}, as implemented in the TRIQS software library~\cite{parcollet_triqs_2015, Priyanka2016cthyb,modest26}. We fix the electronic temperature to $T=1/40$~eV ($\simeq290$~K) unless otherwise noted, above the spin-density wave transition~\cite{chen2024_sdw, chen2024_electronic, kakoi2024_multiband, xie2024_strong}. To mitigate the Monte Carlo sign problem, we perform the calculations in the bonding-antibonding (BA) basis of the bilayer~\cite{ryee2024_quench}, which leads to a diagonal hybridization function.  The bonding ($+$) and antibonding ($-$) combinations of these orbitals are denoted by $x^{\pm}$ and $z^{\pm}$. For more details on the CDMFT, BA basis, and computational settings, see the Supplemental Material (SM).

\textit{Low-energy many-body electronic structure.---}
Figure~\ref{fig:main} presents the calculated low-energy many-body electronic structure of La$_{3}$Ni$_{2}$O$_{7}$ for three representative crystal structures: ambient pressure, $-2\%$ compressive epitaxial strain, and 30~GPa hydrostatic pressure. At ambient pressure, the low-energy spectrum consists of two Fermi surface sheets, $\alpha$ and $\beta$, with mixed $x$- and $z$-orbital characters [see Fig.~\ref{fig:main}(b,c,f)]. A third sheet, labeled $\gamma$ in the DFT calculations (panel b), is shifted below the Fermi level in the many-body theory. The $\alpha$ and $\beta$ sheets have opposite parity under layer exchange. Along the $\Gamma$-M line ($k_{x}=k_{y}$), the $x$ and $z$ orbitals decouple; the relevant band states therefore have purely $x$ orbital character and there is no interlayer hopping. However along the $\Gamma$--X direction, hybridization to the $z$ is symmetry-allowed, and the interlayer coupling  splits the bands so that the $\alpha$ sheet encloses the zone center $\Gamma$, while the $\beta$ sheet surrounds the zone corner M. 

The strongest correlation effects are concentrated within the $z$ orbitals, producing substantial departures from the DFT electronic structure and strongly renormalizing the $z$-derived $\gamma$ band (see Fig.~\ref{fig:main}). Experimentally, the energy position and linewidth of this feature remain contested~\cite{yang2024_arpes,auyeung2025_uni_arpes,wang2025_arpes_films,Li2025nsr}; nevertheless, our ambient-pressure many-body spectrum is in qualitative agreement with available ARPES measurements on bulk single crystals~\cite{yang2024_arpes,auyeung2025_uni_arpes}. We estimate mass renormalizations of $m^{\star}/m_{\mathrm{band}}=5$--$7$ for the $\gamma$ band near M and $m^{\star}/m_{\mathrm{band}}=2$--$3$ for the $\alpha$ and $\beta$ bands near X and $\Gamma$, respectively; see the SM for details.

Hydrostatic pressure and compressive epitaxial strain preserve the basic low-energy band topology, but modify the near-Fermi-level states in qualitatively distinct ways. The $\alpha$ and $\beta$ sheets remain broadly similar to the ambient-pressure case, although the $\beta$ electron pocket is reduced under $-2\%$ epitaxial strain and enhanced at 30~GPa, accompanied by a corresponding change in the pinching of the $\beta$ sheet near X [Fig.~\ref{fig:main}(g,h)]. The largest changes again occur in the $z$-derived $\gamma$ band: under compressive epitaxial strain it shifts to larger binding energy and broadens substantially, indicating reduced coherence and a shorter lifetime, whereas under hydrostatic pressure it moves closer to the Fermi level and retains sharper quasiparticle-like spectral weight. For even higher pressures the band may move through the chemical potential.
Thus, although the $\gamma$ band remains below the Fermi level for experimentally realized pressures in both tuning protocols, its position and coherence are highly sensitive to the crystal structure~\cite{Luo2023,zhang2023structural,liu2023,lechermann2023,qu2024,jiang2024,heier2024,yang2023,sakakibara2024,lu2024interplay,qu2025hund,ouyang2024hund,ji2025strong,yang2023possible, Maier2026Interlayer}.

Our results therefore indicate that hydrostatic pressure and epitaxial strain tune the normal state through distinct evolutions of interlayer correlations, which we will show below is directly correlated to the hybridization between ``dynamical singlets'', formed by out-of-plane interlayer $z$ orbitals, and the itinerant planar $x$ orbitals.

\begin{figure}
    \includegraphics[width=\columnwidth]{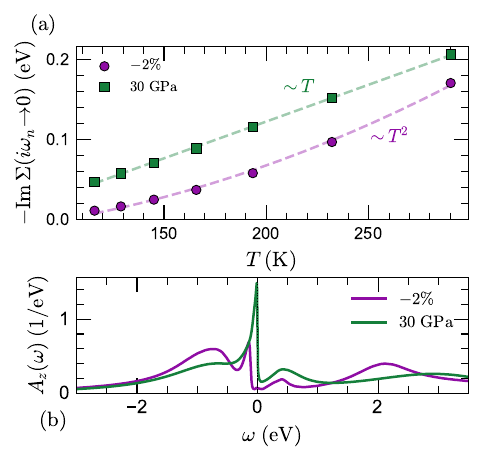}
    \caption{Electron-electron scattering and low-energy spectrum of La$_{3}$Ni$_{2}$O$_{7}$ under hydrostatic pressure (30~GPa) and $-2\%$ compressive strain.
    (a)~Temperature dependence of the scattering rate $\Gamma\propto-\mathrm{Tr}[\mathrm{Im}\Sigma(i\omega_{n}\rightarrow 0)]$.  Dashed lines correspond to fits with forms $f(T) = a + bT^{2}$ and $f(T) = a +bT$ for the strained and pressure case, respectively.
    (b)~Local spectral function for the $z$ orbitals, $A_{z}(\omega)$, at $T=1/100$~eV ($\simeq 116$~K) for 30~GPa (green) and $-2\%$ compressive epitaxial strain (purple).
    }
    \label{fig:fl}
\end{figure}

\textit{Electron-electron scattering and implications for transport.---}Beyond single-particle spectra, transport provides a complementary probe of the normal-state. Recent experiments have shown that pressurized single crystals~\cite{zhang2024strong} and thin-films~\cite{kumar2026nonfermiliquidbehaviorla3ni2o7} of bilayer RP nickelates exhibit an approximately $T$-linear (non-Fermi-liquid) strange-metal resistivity, whereas compressively strained epitaxial thin-films (at ambient pressure) display a more conventional Fermi-liquid $T^{2}$ behavior~\cite{Hsu2026Fermi, kumar2026nonfermiliquidbehaviorla3ni2o7}. To understand this dichotomy, we examine the temperature dependence of the the low-energy scattering rate, $\Gamma\propto-\mathrm{Tr}\left[\mathrm{Im}\Sigma(i\omega_{n}\rightarrow0)\right]$,  for two representative cases: 30~GPa and $-2\%$ compressive epitaxial strain [see Fig.~\ref{fig:fl}(a)]. Assuming, as is typically the case, that the electron scattering rate has the same temperature dependence as the transport scattering rate, our calculations reproduce the experimental trends: the pressurized structure is linear-in-$T$ whereas the strained structure is close to quadratic-in-$T$.

Examination of the orbital-resolved self-energies (see SM) shows that the the $T$-linear behavior arises from fluctuations in the $z$ orbital sector and occurs at the pressure at which the $\gamma$ band reaches the Fermi level, leading to the  sharp quasiparticle feature near the Fermi level shown in Fig.~\ref{fig:fl}(b)]. In contrast, for the $-2\%$ strain case, this feature is shifted to higher binding energy and broadens substantially.

\begin{figure}
    \centering
    \includegraphics[width=\columnwidth]{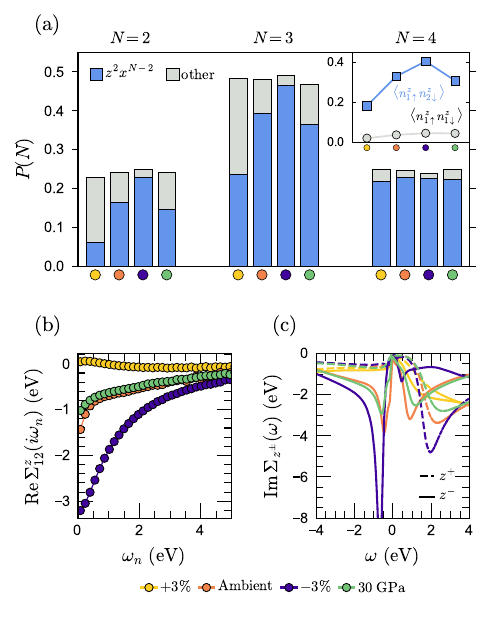}
    \caption{\label{fig:cdmft_trends}
        Signatures of dynamical singlets in La$_{3}$Ni$_{2}$O$_{7}$.
        (a)~Valence histograms of the four orbital quantum impurity problem for different total charge configurations ($N$). For each case (colorful dot), we distinguish with blue color the configurations which have two electrons in the $z$ orbitals ($z^{2}$) and $N-2$ electrons in the $x$ orbitals ($x^{N-2}$) while all other configurations are denoted in grey. Inset: Expectation values for $\langle n_{1\uparrow}^{z}n_{1\downarrow}^{z}\rangle$ (double occupancy) and $\langle n_{1\uparrow}^{z}n_{2\downarrow}^{z}\rangle$ (one $z$ electron on each site with opposite spins.). 
        (b)~Real part of the intersite $z$ orbital self-energy on the Matsubara axis. 
        (c)~Imaginary part of the self-energy for the bonding ($+$) and antibonding ($-$) combinations of the $z$ orbitals on the real frequency axis (after analytic continuation). 
        }
\end{figure}

\textit{Dynamical singlets.---}To identify the microscopic origin of the pressure and strain-dependent electronic structure of La$_{3}$Ni$_{2}$O$_{7}$, we examine how structural tuning reshapes the low-energy physics by changing the non-local electronic correlations.  At the DFT level, the two $z$ orbitals in a bilayer form bonding and antibonding states that are separated in energy by an interlayer hybridization $t_\perp\sim0.5$~eV. Panel (a) of Fig.~\ref{fig:cdmft_trends} shows aspects of the many-body physics of the $z$ orbitals. The total vertical bars show, for different structural cases considered, the probability that the two Ni sites in  unit cell are occupied by two, three or four electrons. We see that the overall occupation probabilities $P(N)$ (sum of blue and grey portions) are very similar for all structural cases. For each occupation probability, we present in blue the probability that there are two electrons in the $z$ orbitals (with the remaining particles if any in the $x$ orbitals).  Across the charge sectors, the dominant weight lies in the two-electron $z$-orbital subspace ($z^{2}x^{N-2}$) indicating a dominance of the $z$-dimer configuration, except in the case of $3\%$ tensile strain where the probabilities are more uniformly distributed. As epitaxial compression is increased, the dominance of the dimer configuration becomes more pronounced. By $-3\%$, the electronic configuration is essentially completely a $z$ dimer with the remaining electron in the $x$ orbitals.

The inset probes the physics of these dimer configurations. The grey circles show the double occupation of a single $z$ orbital ($\langle n_{1\uparrow}^{z}n_{1\downarrow}^{z}\rangle$), which remains small in all cases. The blue squares show the probability that the configuration is of one electron on each $z$ orbital, with opposite spin -- in other words, a dimer singlet. We see that the probability is maximized at large compressive epitaxial strain, where the singlet is essentially the only configuration.  This is reminiscent of an orbital-selective Mott state, but with an important distinction: the two electrons on neighboring $z$ orbitals are predominantly locked into a spin singlet.

Panels (b) and (c) of Fig.~\ref{fig:cdmft_trends} show two representations of the self-energy of the two $z$ orbitals in the bilayer. Figure~\ref{fig:cdmft_trends}(b) presents the intersite component $\mathrm{Re}\Sigma^{z}_{12}(i\omega_{n})$  as a function of Matsubara frequency for different strain and pressure states.  As the epitaxial strain is increased from $3\%$ tensile to $3\%$ compressive, the intersite self-energy goes from being small in value and weakly frequency-dependent to large and strongly frequency-dependent, a signature of intersite correlations increasing with compressive strain. Figure~\ref{fig:cdmft_trends}(c) shows the analytically continued $z$ self-energy in the bonding-antibonding basis, revealing two pole-like structures at $\omega\approx -1$~eV and $\approx 2$~eV, which grow in strength as the compressive strain is increased.

\begin{figure}
    \centering
    \includegraphics[width=\columnwidth]{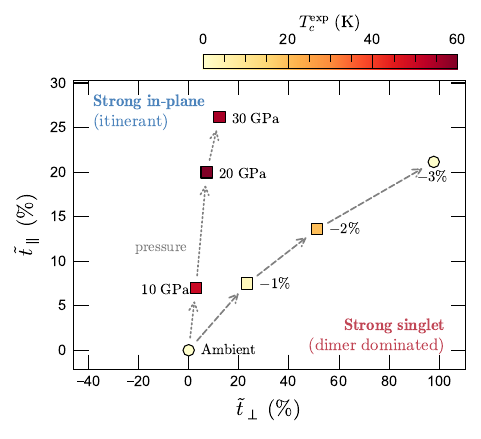}
    \caption{\label{fig:cdmft_diagram}~Effect of pressure on the low-energy effective in-plane ($\tilde t_\parallel$) and out-of-plane ($\tilde t_\perp$) hoppings of the bilayer RP nickelates. Definitions in the main text. The values are expressed as percent change from the ambient pressure calculation. The marker color corresponds to the experimentally observed superconducting transition temperature ($T^{\text{exp}}_{c}$) shown on the colorbar. 
    }
\end{figure}

To interpret these results, we recall that, in the Hubbard dimer, the imaginary part of the self-energy in the bonding-antibonding basis is $\frac{\pi U^2}{4}\delta(\omega\mp 3t)$, corresponding to $\text{Re}\Sigma_{12}(i\omega_n)=\frac{U^2}{4}\frac{3t}{\omega_n^2+9t^2}$. At large compressive strain, $\text{Re}\Sigma_{12}(i\omega_n)$ is approximately of this form with $3t\sim 1.5$~eV and $U\approx 3$~eV. Correspondingly, the imaginary part of the real-axis self-energy in the bonding-antibonding basis displays poles at $\omega\approx 2$~eV and $\approx -0.8$~eV respectively. As the compressive strain is reduced, the pole amplitude decreases and additional structures arise from the hybridization with the planar $x$ orbitals.

\textit{Interlayer versus intralayer correlations.---}A different perspective on the preceding results is to summarize the normal-state of La$_{3}$Ni$_{2}$O$_{7}$ in terms of two renormalized low-energy scales: an effective intralayer $x$ hopping, $\tilde{t}_{\parallel}$, and an effective interlayer $z$ hopping, $\tilde{t}_{\perp}$. We define these quantities as $\tilde{t}_{\parallel} \equiv Z_{x}t_{\parallel}$ and $\tilde{t}_{\perp} \equiv Z_{z}[t_{\perp} + \mathrm{Re}\Sigma_{12}^{z}(0)]$, where $t_{\parallel}$ and $t_{\perp}$ are the bare intra- and interlayer hoppings, respectively.  In this parameterization, $\tilde{t}_{\parallel}$ tracks the degree of in-plane itinerancy, while $\tilde{t}_{\perp}$ tracks the strength of the interlayer $z$ orbital and its tendency toward singlet formation.

Figure~\ref{fig:cdmft_diagram} shows that hydrostatic pressure and compressive epitaxial strain reorganize $\tilde{t}_{\perp}$ and $\tilde{t}_{\parallel}$ in qualitatively different ways. Non-local correlations strongly enhance $\tilde{t}_{\perp}$ through $\Sigma_{12}^{z}$, with the largest effect under compressive strain. In these structures, the enlarged $c$-axis and reduced intercell hopping isolate the Ni bilayers within each perovskite block, amplifying the effective intra-dimer $z$ orbital coupling despite only small changes in the Ni-O-Ni bond lengths. Hydrostatic pressure, by contrast, modifies $\tilde{t}_{\perp}$ only moderately, but substantially increases $\tilde{t}_{\parallel}$ by broadening the in-plane bandwidth and weakening correlations in the $x$ orbitals. Pressure and strain therefore move the system along different directions in the low-energy phase space: strain predominately strengthens the interlayer correlations, whereas pressure primarily enhances in-plane itinerancy.

Overlaying the experimental superconducting transition temperature $T_{c}$ further highlights this contrast. One interpretation is that hydrostatic pressure promotes superconductivity mainly through an increase of the intralyer exchange ($\propto \tilde{t}_{\parallel}^{2}$). Compressive strain also enhances the in-plane superexchange, but simultaneously drives the system toward a strong singlet regime that shifts spectral weight into dynamical singlets. Within this picture, how superconductivity in the bilayer RP nickelates emerges from this dynamical singlet normal-state warrants future theoretical investigations~\cite{khaliullin2026orbital}.

\textit{Conclusions.---}Using CDMFT, we have shown that local and interlayer electronic correlations reshape the normal-state electronic structure of bilayer RP nickelates in qualitatively different ways under hydrostatic pressure and epitaxial strain. Our results account for the contrasting photoemission and transport signatures of strained thin-films and bulk single crystals, and identify a crucial aspect of the low-energy state as interlayer dynamical singlets built out of the $z$ orbitals, whose hybridization with the planar $x$ orbitals gets weaker as the compressive strain increases. Our picture may be thought of as a variant of the ``orbital-selective Mott insulator'' concept~\cite{Anisimov02,Koga04,Medici09}, in which one of the orbitals (in our case the $z$ orbital) becomes Mott localized, hosting a magnetic moment, while other orbitals (here the $x$ orbitals) remain itinerant but may be Kondo-coupled to the local moments. In the RP nickelates, this concept is extended because the localized orbitals form dynamical singlets, decoupling from the itinerant bands. This picture provides a natural starting point for future theoretical studies of the electronic instabilities of this low-energy theory, such as spin-density waves and superconductivity, and for experiments designed to probe and manipulate this novel normal-state. In particular, as suggested by Khaliullin and Chaloupka~\cite{khaliullin2026orbital} and, relatedly, by Maier \textit{et al.} ~\cite{Maier2026Interlayer}, pairing via virtual scattering into the excited triplet state of the $z$ dimers is an intriguing possibility. More generally, the model realized by this family of materials provides an interesting laboratory for studying the orbital-selective Mott transition and its interplay with intersite ordering and correlations.

\begin{acknowledgments}
\textit{Acknowledgments.---}We thank A. S. Botana, V. Pardo, M. R. Norman, and J. F. Mitchell for many fruitful discussions and A. Devarakonda for a careful reading of the manuscript. The Flatiron Institute is a division of the Simons Foundation.
\end{acknowledgments}

\bibliography{ref.bib}

\end{document}


\title{Supplemental Material:\\Squeezing dynamical singlets in bilayer nickelates
}
\author{Harrison LaBollita\,\orcidlink{0000-0002-6699-8577}}
\email{hlabollita@flatironinstitute.org}
\affiliation{Center for Computational Quantum Physics, Flatiron Institute, New York, New York 10010, USA}
\author{Andrew J. Millis\,\orcidlink{0000-0001-9413-6344}}
\affiliation{Center for Computational Quantum Physics, Flatiron Institute, New York, New York 10010, USA}
\affiliation{Department of Physics, Columbia University, New York, New York 10027}
\author{Olivier~Gingras\,\orcidlink{0000-0003-3970-6273}}
\affiliation{Center for Computational Quantum Physics, Flatiron Institute, New York, New York 10010, USA}
\affiliation{Université Paris-Saclay, CNRS, CEA, Institut de physique théorique, 91191, Gif-sur-Yvette, France}

\maketitle

\section{\label{supp:dft}Density-functional theory and Crystal structures}
Density-functional theory (DFT) calculations were performed with Q\textsc{uantum} ESPRESSO~\cite{qe2009,qe2017} using norm-conserving pseudopotentials parameterized for the Perdew-Burke-Ernzerhof (PBE) exchange correlation functional~\cite{Perdew:1996iq}. The plane-wave basis set size was set with a kinetic energy cutoff of 100 (400)~Ry for the wavefunctions (charge density). Brillouin zone integration was performed on a $12\times12\times12$ ($7\times7\times7$) $\textbf{k}$-grid with a Methefessel-Paxton smearing of 0.005~Ry for self-consistent field (structural relaxation) calculations.

The crystal structure was initialized in the body-centered tetragonal ($I4/mmm$) phase of La$_{3}$Ni$_{2}$O$_{7}$ using lattice parameters from Ref.~\citenum{zhao2025} for both ambient-pressure and epitaxial strained configurations. For the strained structures, internal atomic coordinates were fully relaxed while keeping lattice parameters fixed. Hydrostatic pressure was simulated by relaxing both lattice parameters and internal coordinates under an applied external stress tensor. The resulting structural parameters are summarized in Table~\ref{tab:structures}. Other than the obvious changes of the in-plane and out-of-plane lattice constant corresponding to the external perturbations (either hydrostatic pressure or epitaxial strain), we find that the terminal apical oxygen distance changes most dramatically relative to the other bonds. Because epitaxial strain controls the in-plane lattice constant, while the out-of-plane responds via Poisson's ratio, this perturbation dramatically shifts the outer terminal apical bonds.

\begin{table}[h!]
    \centering
    \caption{
        \label{tab:structures}
        Selected crystal structure data for La$_{3}$Ni$_{2}$O$_{7}$ ($I4/mmm$) for different epitaxial strains and hydrostatic pressures. The Ni-O bond distances are: basal (in-plane), bridging apical (Ni-O bond distance between perovskite layers), and terminal apical (Ni-O bond distance point towards the rocksalt spacer layer). Structural data is organized from smallest in-plane ($a$) lattice constant to largest. All data is in units of \AA.
    }
    \begin{tabular*}{\linewidth}{@{\extracolsep{\fill}}r|ccccc}
        \toprule
        Pressure/strain & $a$ & $c$ & Ni-O (basal) & Ni-O (bridging apical) & Ni-O (terminal apical)\\
        \hline
        30 GPa        & 3.680  & 19.41    &  1.84  &  1.90    &  2.05  \\
        20 GPa        & 3.725  & 19.66    &  1.86  &  1.92    &  2.09  \\
        10 GPa        & 3.778  & 19.97    &  1.89  &  1.94    &  2.14  \\
        $-3 \%$       & 3.730  & 21.14    &  1.87  &  1.97    &  2.38  \\
        $-2 \%$         & 3.769  & 20.87    &  1.88  &  1.97    &  2.32  \\
        $-1 \%$        & 3.807  & 20.63    &  1.90  &  1.96    &  2.26  \\
        Ambient       & 3.847  & 20.40    &  1.92  &  1.96    &  2.22  \\
        $+3 \%$        & 3.962  & 19.81    &  1.98  &  1.95    &  2.08  \\
        \bottomrule
    \end{tabular*}
\end{table}

\begin{figure}
    \includegraphics[width=\columnwidth]{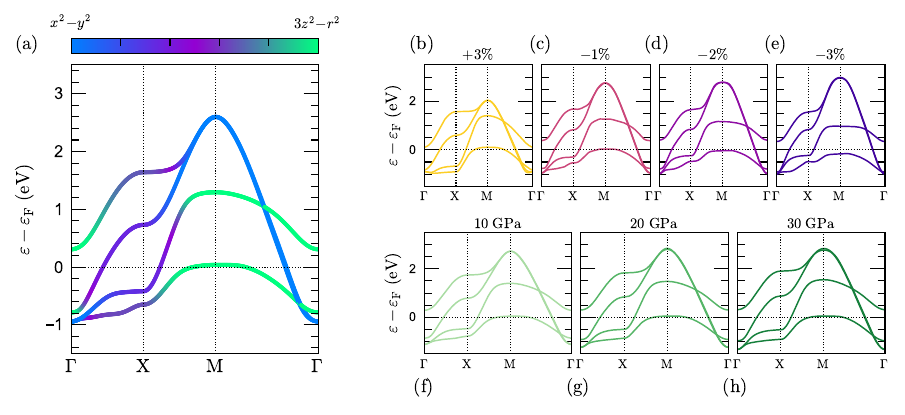}
    \caption{
        \label{fig:bands}
        Wannier dispersions representing the low-energy electronic structure of La$_{3}$Ni$_{2}$O$_{7}$ obtained based on DFT. 
        (a)~Wannier dispersion for the antibonding Ni-$e_{g}$ bands of La$_{3}$Ni$_{2}$O$_{7}$ at ambient pressure. Color denotes the orbital content of the state at ($\mathbf{k}$, $\varepsilon_{\nu}(\mathbf{k})$) where Ni-$d_{x^{2}-y^{2}}$ corresponds to blue and Ni-$d_{z^{2}}$ corresponds to green. Mixed orbital character is shown in purple.
        (b-h)~Wannier dispersion for La$_{3}$Ni$_{2}$O$_{7}$ for different pressure and epitaxial strains.
        (b-e)~Epitaxial strain: +3\%, -1\%, -2\%, and -3\% (left to right).
        (f-h)~Hydrostatic pressure: 10 GPa, 20 GPa, and 30 GPa (left to right)
    }
\end{figure}


\section{\label{supp:w90}Low-energy electronic structure}
For each structural geometry, we obtain a Wannier representation of the near-Fermi level states with Ni-$e_{g}$ character using maximally-localized Wannier functions as implemented in Wannier90~\cite{wannier90, Pizzi_2020}. This results in four-orbital (two $e_{g}$ orbitals per layer) that represents the one-body Hamiltonian in the following orbital basis: $\{d_{1,z^{2}}, d_{1,x^{2}-y^{2}}, d_{2,z^{2}}, d_{2,x^{2}-y^{2}}\}$ where $z^2$ denotes the $3z^2-r^2$ orbital. The low-energy electronic structure for each of these crystal structures is summarized in Fig.~\ref{fig:bands}. The near-Fermi level bandstructure corresponds to antibonding bands constructed from hybridized transition-metal (Ni) and ligand (O) states. We show the low-energy Wannier dispersions along high-symmetry lines in the tetragonal Brillouin zone at ambient pressure (see Fig.~\ref{fig:bands}(a)). The bandwidth ($W$) is roughly $\sim 3.5$~eV. The four bands have mixed Ni-$e_{g}$ character. At $\Gamma$, the states have pure orbital character. The $d_{x^{2}-y^{2}}$ states (blue) are nearly degenerate, which reflects their lack of interlayer coupling, while the $d_{z^2}$ states show a large splitting indicating their significant hybridization via the bridging apical oxygen. Moving along the zone from $\Gamma$ to X, the bands are dispersive with significant orbital admixture (purple). Along the zone diagonal ($\Gamma$ to M), we find that the band structure retains pure orbital character with flat $d_{z^2}$ (green) and dispersive $d_{x^{2}-y^{2}}$ (blue) bands. The evolution of these basic dispersions is summarized in Figs~\ref{fig:bands}(b-h) under epitaxial strain (see Figs~\ref{fig:bands}(b-e)) and hydrostatic pressure (see Figs~\ref{fig:bands}(f-h)).


\section{Cluster dynamical mean-field theory}

\subsection{Notation and setup}
We consider the four-orbital Wannier model as our low-energy representation for the electronic structure of La$_{3}$Ni$_{2}$O$_{7}$, described by $H = H_{0} + H_{\mathrm{int}}$. The non-interacting part is
\begin{equation}
    H_{0}(\mathbf{R}) = \sum_{ij,lm,\sigma}t_{il,jm}(\mathbf{R})d^{\dagger}_{il\sigma}d_{jm\sigma},
\end{equation}
where $l,m\in \{z^{2},x^{2}-y^{2}\}$ label the Ni-$e_{g}$ orbitals and $i,j$ label the two layers. The hopping amplitudes $t_{il,jm}(\mathbf{R})$ are the real-space hopping integrals obtained from the Wannier fits to the DFT data as described in Sec.~\ref{supp:dft}. The interaction term is given by the onsite Hubbard-Kanamori Hamiltonian,
\begin{equation}
    H_{\mathrm{int}}=\frac{1}{2}\sum_{ijkl,\sigma\sigma^{\prime}}U_{ijkl}d^{\dagger}_{i\sigma}d^{\dagger}_{j\sigma^{\prime}}d_{l\sigma^{\prime}}d_{k\sigma},
\end{equation}
where $U_{ijkl}$ is $U_{mmmm} = U$, $U_{mm'mm'} = U^{\prime}$, $U_{m m'm'm} = U_{mmm'm'} = J$. We choose $U = 4$~eV, $J = 0.8$~eV, and $U^{\prime}= U - 2J = 2.4$~eV, in line with previous works~\cite{christiansson2023,ryee2024_quench}. We only consider local interactions on each Ni site, respectively. To mitigate the Monte Carlo sign problem, the model is solved in the bonding-antibonding basis defined by symmetric and antisymmetric combinations of the two layers:
\begin{equation}
    d_{\eta^{\pm}\sigma} = \frac{1}{\sqrt{2}}(d_{1,\eta,\sigma} \pm d_{2,\eta,\sigma}),
\end{equation}
where $\eta^\pm$ represents the bonding ($+$) and antibonding ($-$) combinations of orbital $\eta \in \{z^{2}, x^{2}-y^{2}\}$. Both $H_{0}$ and $H_{\mathrm{int}}$ are transformed accordingly to this bonding-antibonding basis. We solve the quantum impurity model self-consistently with a fixed target density of $n = 3$ electrons, corresponding to an average filling of $d^{7.5}$ (Ni$^{2.5+}$) based on the formal valence counting of La$_{3}$Ni$_{2}$O$_{7}$ (La$^{3+}$, Ni$^{2.5+}$, O$^{2-}$).

\subsection{Computational details}
The CDMFT loop itself is written in using the TRIQS/ModEST library~\cite{modest26}. We solve the four-orbital quantum impurity problem corresponding to the Ni-$e_{g}$ cluster using the continuous-time quantum Monte Carlo (CTHYB) solver within the hybridization expansion~\cite{PhysRevLett.97.076405, RevModPhys.83.349} as implemented in the TRIQS software library~\cite{Parcollet2015triqs,Seth2016274}. We solve the impurity problem at an electronic temperature $T = 1/40$~eV ($\simeq290$~K) for all crystal structures. For the crystal structures corresponding to $-2\%$ and 30~GPa, we sweep a range of temperatures from $T = 1/100$--$1/40$~eV ($\simeq116-290$ K). To obtain the high-frequency tail of the self-energy, we fit the tail to its high-frequency expansion enforcing the first two moments based on their analytic expression determined by the measured impurity density matrix~\cite{labollita25}. We chose $(\omega_{\mathrm{min}},\omega_{\mathrm{max}}) = (10, 14)$~eV fo the tail-fitting window. To obtain real-frequency data, we employ the maximum entropy method of analytic continuation as implemented in TRIQS/Maxent~\cite{Kraberger2017maxent}.

\section{Many-body renormalization of bare Wannier Hamiltonian}

\begin{figure}
    \centering
    \includegraphics[width=\columnwidth]{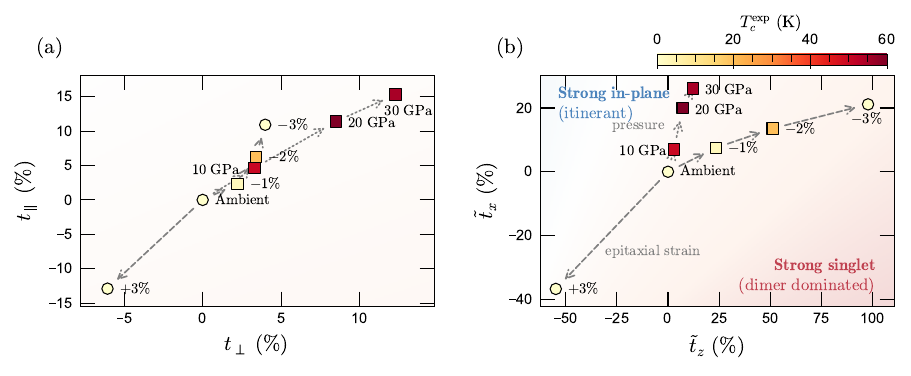}
    \caption{
        \label{fig:renorm}
        Renormalization of the intralayer and interlayer hopping integrals due to electronic interactions in La$_{3}$Ni$_{2}$O$_{7}$. 
        (a)~Scatter plot of interlayer ($t_{\perp}$) versus intralayer ($t_{\parallel}$) hopping extracted from Wannier Hamiltonian fits to DFT band structures under hydrostatic pressure and epitaxial strain. Here $t_{\perp}$ denotes hopping between $z$ orbitals within a bilayer, while $t_{\parallel}$ corresponds to the nearest-neighbor in-plane hopping.
        (b)~Same as (a), but for interaction-renormalized hopping parameters (see also Fig.~1(f) in the main text). All values are shown as percentage changes relative to ambient pressure.
    }
\end{figure}

Figure~\ref{fig:renorm} illustrates the many-body renormalization of the bare Wannier hopping integrals. A $4\times4$ Wannier Hamiltonian is constructed by fitting the near Fermi-level DFT band structures for crystal structures obtained under hydrostatic pressure and epitaxial strain. The resulting nearest-neighbor intralayer hopping ($t_{\parallel}$) and interlayer hopping ($t_{\perp}$) are shown in Fig.~\ref{fig:renorm}(a).

Under hydrostatic pressure, both $t_{\parallel}$ and $t_{\perp}$ increase relative to their ambient pressure values. Epitaxial strain produces a similar trend for $t_{\parallel}$, but leads to a significantly weaker enhancement of $t_{\perp}$. This difference has been proposed as a possible explanation for the reduced $T_{c}$ observed in epitaxial strained thin films, assuming superconductivity is mediated by interlayer spin fluctuations~\cite{li2026_enhanced}. However, as discussed in the main text, the inclusion of local and non-local electronic correlations qualitatively alters this picture. Figure~\ref{fig:renorm}(b) reproduces the same scatter plot, but now using renormalized hopping parameters. With many-body effects included, a clear separation emerges between hydrostatic pressure and epitaxial strain, highlighting the importance of correlation-driven renormalizations.

\section{Comparison with single-site DMFT}
For the ambient pressure crystal structure, we compare cluster DMFT (CDMFT) and single-site DMFT calculations. In the single-site DMFT approximation, the two Ni layers are decoupled, resulting in two equivalent two-orbital quantum impurity problems. In contrast, the CDMFT calculation treats the bilayer explicitly as a four-orbital impurity problem, thereby incorporating short-range intersite correlations.

\begin{figure}
    \centering
    \includegraphics[width=0.9\columnwidth]{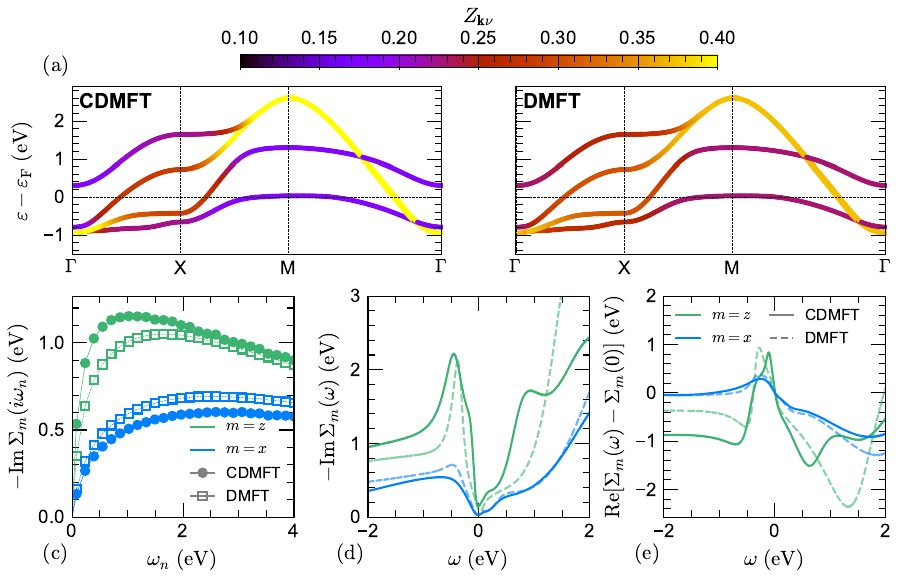}
    \caption{
        \label{fig:comparison}
        Comparison of self-energies from CDMFT and single-site DMFT for La$_{3}$Ni$_{2}$O$_{7}$ at ambient pressure. 
        (a,b)~Momentum-resolved quasiparticle weight in the band basis $Z_{\vec{k}\nu}$ projected onto the Wannier band structure along high-symmetry line for (a) CDMFT and (b) DMFT. The colorscale is identical in both panels.
        (c)~Imaginary part of the diagonal (local) Matsubara self-energy.
        (d)~$-\mathrm{Im}\Sigma(\omega)$ on the real-frequency axis.
        (e) $\mathrm{Re}[\Sigma(\omega)-\Sigma(0)]$, orbitally-resolved for CDMFT (solid) and DMFT (dashed). 
    }
\end{figure}

A comparison of the resulting self-energies and quasiparticle renormalizations is presented in Fig.~\ref{fig:comparison}. Panels (a) and (b) of Figure~\ref{fig:comparison} show the momentum-resolved quasiparticle weight $Z_{\vec{k}\nu}$ projected onto the Wannier band structure along high-symmetry directions for CDMFT and DMFT, respectively (both panels share the same colorbar). $Z_{\vec{k}\nu}$ is obtained by first transforming the Matsubara self-energy into the band basis,
\begin{equation}
\label{eqn:bands_sig}
    \Sigma_{\nu\nu'}(\vec{k},\omega) = \sum_{mm'}\langle \psi_{\vec{k}\nu}|\chi_{m\vec{k}}\rangle\Sigma_{mm'}(\omega)\langle\chi_{m'\vec{k}}|\psi_{\vec{k}\nu'}\rangle,
\end{equation}
then evaluating the standard quasiparticle weight from the low-frequency derivative of the imaginary part of the self-energy. The CDMFT results exhibit a strong orbital differentiation between the planar ($x$) and the interlayer ($z$) orbitals, in better agreement with ARPES measurements~\cite{yang2024_arpes}. While single-site DMFT captures the same qualitative trend, the effect is significantly weaker. This distinction is further evident in the orbital-resolved self-energies on both Matsubara and real-frequency axes, shown in Fig.~\ref{fig:comparison}(c-e).

\begin{table}
\centering
\caption{\label{tab:mass}Band-resolved mass enhancements ($m^{\star}/m_{\mathrm{band}}$) at high-symmetry points from CDMFT at ambient and $-2\%$ strain compared with experimental values taken from ARPES on bulk single-crystals~\cite{yang2024_arpes} and thin-films~\cite{wang2025_arpes_films}.} 
\begin{tabular*}{\linewidth}{@{\extracolsep{\fill}}cccccc}
\toprule
$k$-point & Band & Ambient (CDMFT) & Bulk~\cite{yang2024_arpes} & $-2\%$ (CDMFT) & Thin-films~\cite{wang2025_arpes_films}\\
\hline
X & $\gamma$ & 4.5 & 5.5 $\pm$ 1 & 5.7  & 7$\pm$3\\
  & $\alpha$ & 3.4 & 1.8 $\pm$ 1 & 3.9  & 2.9$\pm$0.9 \\
  & $\beta$  & 3.5 & 2.6 $\pm$ 1 & 3.1  & 3.8$\pm$0.9\\
\hline
M & $\gamma$       & 5.6 & 8.3 $\pm$ 1 & 6.5 & --\\
  & $\alpha/\beta$ & 2.4 & 2.3 $\pm$ 1 & 2.3 & --\\
\bottomrule
\end{tabular*}
\end{table}

\section{Additional data}

\subsection{Matsubara self-energies}

\begin{figure}
    \centering
    \includegraphics[width=\columnwidth]{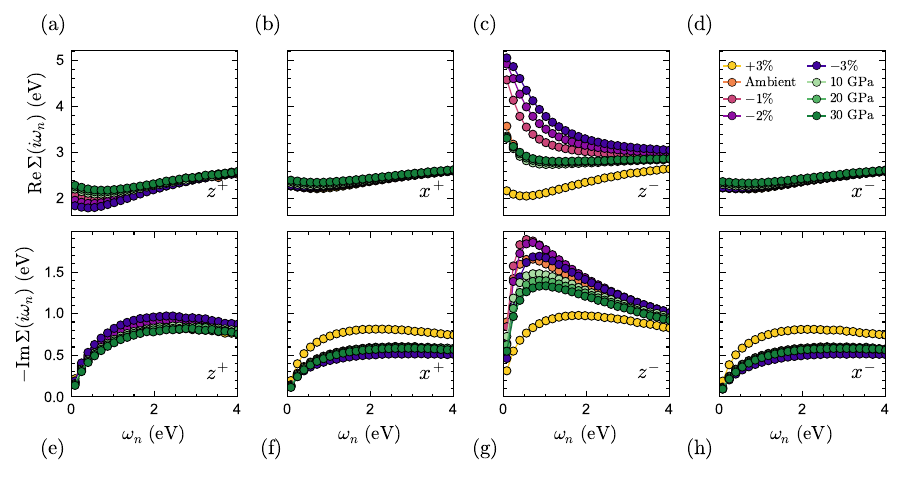}
    \caption{Matsubara self-energies from CDMFT in the bonding-antibonding basis for all structural cases of La$_{3}$Ni$_{2}$O$_{7}$. (a)-(d) $\mathrm{Re}\Sigma_m(i\omega_{n})$. (e-h) $\mathrm{Im}\Sigma_m(i\omega_{n})$.}
    \label{fig:sig_iw_ab}
\end{figure}

Figure~\ref{fig:sig_iw_ab} shows the Matsubara self-energies of our four-orbital quantum impurity model obtained from CTHYB at an inverse system temperature of $\beta = 40$ eV$^{-1}$, where $z^{\pm}$ and $x^{\pm}$ are the bonding ($+$) and antibonding ($-$) combinations of the $z$ and $x$ orbitals, respectively. As expected $\Sigma_{x^{+}}(i\omega_{n}) \approx \Sigma_{x^{-}}(i\omega_{n})$ for all cases indicating the weak hybridization between the planar $x$ orbitals on the two layers. However, there is clear contrast between $\Sigma_{z^{+}}(i\omega_{n})$ and $\Sigma_{z^{-}}(i\omega_{n})$ which is indicative of the large hybridization between the $z$ orbitals on the two layers. At low-frequencies, the magnitude of $\mathrm{Re}\Sigma^{z^{\pm}}$ results in the many-body correction to the DFT on-site energy levels.

\begin{figure}
\centering
\includegraphics[width=\columnwidth]{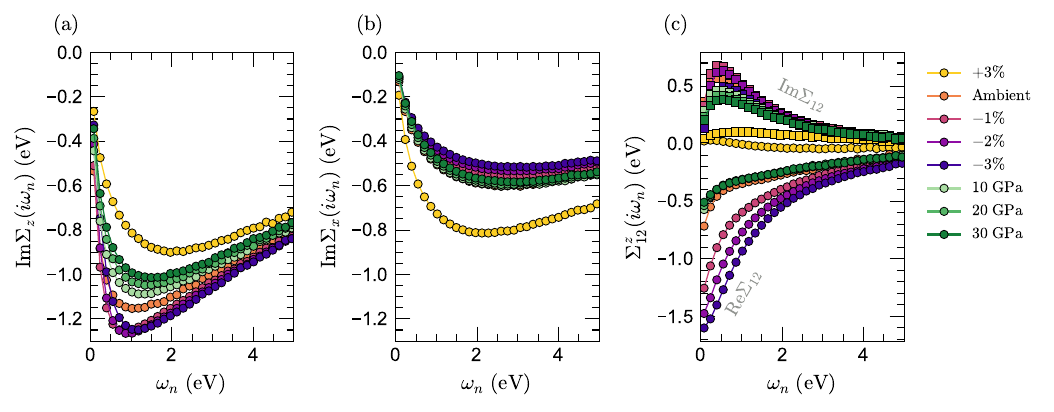}
\caption{\label{fig:sigma_iw}Matsubara self-energy from CDMFT of La$_{3}$Ni$_{2}$O$_{7}$ under external perturbations (hyrdostatic pressure and epitaxial strain). (a) Imaginary part of the Matsubara self-energy for the (a) $z$ and (b) $x$ orbitals. (c) Real and imaginary part of the intersite self-energy between the $z$ orbitals (square and circle markers denote the imaginary and real parts, respectively).}
\end{figure}

From the imaginary part of the Matsubara self-energies, we find that the correlated physics mainly involves the $z^{\pm}$ orbitals with $|\mathrm{Im}\Sigma_{z^{\pm}}(i\omega_{n})| > |\mathrm{Im}\Sigma_{x^{\pm}}(i\omega_{n})|$. We observe clear trends starting from the 3\% tensile case to the 3\% compressive case, which clearly tunes the interlayer correlations between the $z$ orbitals, while the $x$ orbitals remain similarly moderately correlated across all cases.

To complement Fig.~\ref{fig:sig_iw_ab}, we transform the Matsubara self-energies to the site basis. Figure~\ref{fig:sigma_iw} shows the orbital-resolved local and non-local components of the Matsubara self-energies obtained from CDMFT for different external perturbations (hydrostatic pressure and epitaxial strain). The on-site local components for both orbitals is linear at low-frequency: $\Sigma_{\eta}^{\mathrm{local}} = \Sigma_{\eta}^{\mathrm{local}}(0) + (1-1/Z_{\eta})\omega + \cdots$. As pointed out in the main text, the significant increase in the interlayer correlations under epitaxial compressive strain can be understood as the gapping out of the low-energy quasiparticles associated with the interlayer dynamical singlets~\cite{biermann2005}.

For comparison with experiment, we tabulate the mass renormalization for ambient-pressure and $-2\%$ compressive epitaxial strain using Eq.~\eqref{eqn:bands_sig} in Table~\ref{tab:mass}. We find excellent qualitative agreement between the calculated mass renormalizations and the available experimental along two different momentum cuts in the Brillouin zone.

\begin{figure*}
    \centering
    \includegraphics[width=0.75\columnwidth]{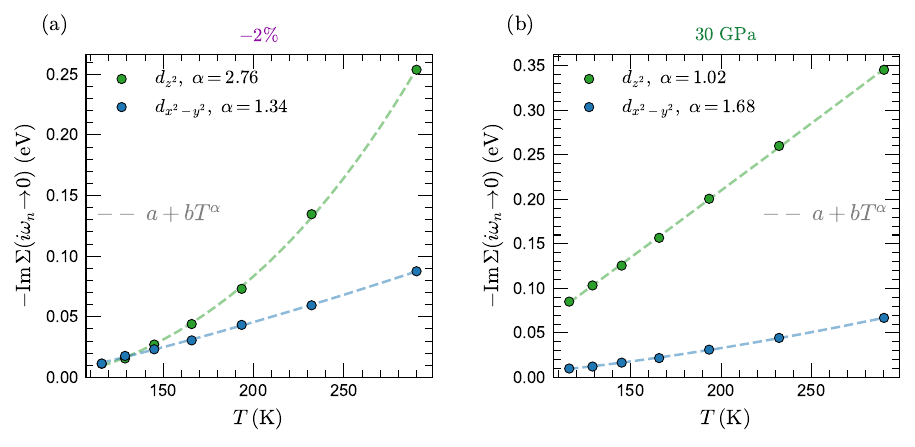}
    \caption{Orbital-resolved scattering rates ($-\mathrm{Im}\Sigma_{m}(i\omega_{n}\rightarrow 0)$ for La$_{3}$Ni$_{2}$O$_{7}$ for two different structural cases: (a) $-2\%$ compressive epitaxial strain and (b) 30 GPa. The dashed lines denote fits to the form $a + bT^{\alpha}$. The power-law of $T$ is denoted in the legend for each orbital.}
    \label{fig:orb_scatt}
\end{figure*}

The temperature dependence of the low-energy scattering rate is directly related to the transport properties of a materials within the dynamical mean-field approximation. Panels (a) and (b) Fig.~\ref{fig:orb_scatt} show the orbital-resolved scattering rate for the $-2\%$ strain and 30 GPa crystal structures, respectively. The dashed lines in Fig.~\ref{fig:orb_scatt} corresponds to fitting the data to the form $a + bT^{\alpha}$, where $\alpha$ corresponds to the power-law of the scattering rate. For a Fermi liquid, $\alpha = 2$, while for a non-Fermi-liquid strange metal, $\alpha = 1$. Comparing the two structural cases together, we observe that the $z$ component of the self-energy drives the change in power-law, which as we discuss in the main text is related to the location of the quasiparticle peak in the $z$ orbital spectral function.

\subsection{Real-frequency self-energies}
\begin{figure}
    \centering
    \includegraphics[width=\columnwidth]{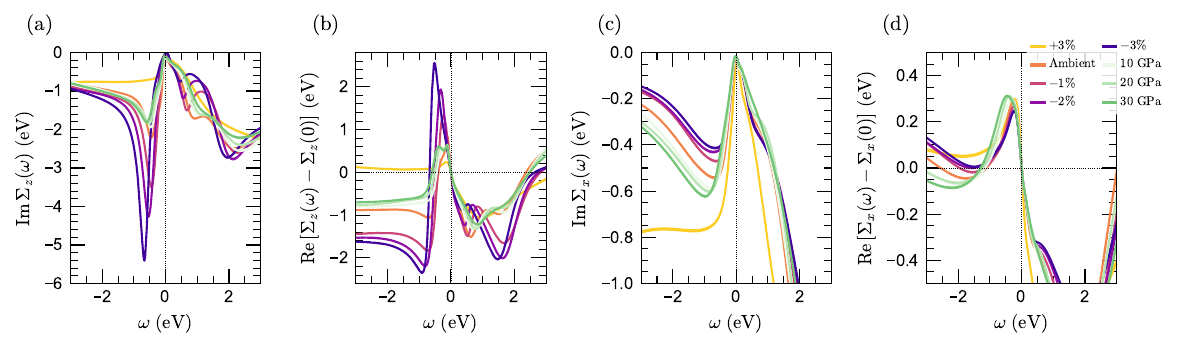}
    \caption{Orbital-resolved self-energy in real frequencies (via analytic continuation) for each structural case considered for La$_{3}$Ni$_{2}$O$_{7}$. $\Sigma_{z}(\omega)$ for the (a) $\mathrm{Im}\Sigma_{z}(\omega)$ and (b) $\mathrm{Re}\Sigma_{z}(\omega)$. $\Sigma_{x}(\omega)$ for (c) $\mathrm{Im}\Sigma_{x}(\omega)$ and (d) $\mathrm{Re}\Sigma_{x}(\omega)$.}
    \label{fig:sig_w}
\end{figure}

Figure~\ref{fig:sig_w} shows the orbital-resolved self-energies for each orbital after analytic continuation (via the maximum entropy method). Panels (a) and (b) of Fig.~\ref{fig:sig_w} correspond to the imaginary and real parts of the $z$ orbital self-energies, respectively. Starting from the 3\% compressive epitaxial strain case, we observe approximate poles in the imaginary part of the real-frequency self-energy near $\omega \approx 2$ eV and $\omega \approx -0.5$ eV. As compressive strain is reduced, an additional low-energy structure between the poles around $\omega \approx 0.5$ eV arises and corresponds to the hybridization with the planar $x$ orbitals. 

\bibliography{ref}